\documentstyle[12pt,epsfig]{article}
     \def\lsim{\raise0.3ex\hbox{$<$\kern-0.75em\raise-1.1ex\hbox{$\sim$}}}
\def\gsim{\raise0.3ex\hbox{$>$\kern-0.75em\raise-1.1ex\hbox{$\sim$}}}
\def\noi{\noindent}

\def\bea{\begin{eqnarray}}  \def\eea{\end{eqnarray}}
\def\beq{\begin{equation}}   \def\eeq{\end{equation}}

\def\beeq{\begin{eqnarray}} \def\eeeq{\end{eqnarray}}

\begin{document}
\begin{center}
\vbox to 1 truecm {}
{\Large \bf }\par 
\vskip 3 truemm
  {\Large \bf Radial Flow in a Final State Interaction Model}  \vskip 1.5 truecm
{\bf A. Capella}\\ 
Laboratoire de Physique Th\'eorique\footnote{Unit\'e Mixte de
Recherche UMR n$^{\circ}$ 8627 - CNRS}
\\ Universit\'e de Paris XI, B\^atiment 210, 91405 Orsay Cedex, France
\vskip 5 truemm
{\bf E. G. Ferreiro}\\
Departamento de F{\'\i}sica de Part{\'\i}culas, 
Universidad de Santiago de Compostela, 15782 Santiago de Compostela, 
Spain

\end{center}
\vskip 1 truecm
\begin{abstract}
In the framework of a final state interaction model, we show that the so-called radial flow, i.e. the almost linear increase of the inverse slope $T$ with the mass of the produced particle, is already contained in the initial condition -- with a slope $<u_t^2>$ (the so-called strength of the average radial transverse flow) which is larger than the measured one. While the precise value of the slope depends on the details of the model, the above result has a very general basis -- namely the increase with increasing $p_T$ of the fixed $p_T$ suppression, in the low $p_T$ region. 

\end{abstract}

\vskip 2 truecm
\noi LPT-Orsay-06-33\par
\noi May 2006
\newpage
\pagestyle{plain} 

Flows of different types are considered as providing strong support to hydrodynamic models and, more generally, as signals of the production of a system in thermal equilibrium in heavy ion collisions. In particular the agreement of the hydrodynamic model predictions on elliptic flow with the experimental data at RHIC energies constitutes the basis of the claim that a perfect fluid has been produced at RHIC \cite{1r}. This interpretation has been challenged in \cite{2r} where the authors claim that the RHIC data on elliptic flow are inconsistent within the hydrodynamic model predictions for a perfect fluid (i.e. with no viscosity). Moreover, an alternative description of elliptic flow at RHIC has been proposed \cite{3r} in the framework of a final state interaction model -- similar to the one previously introduced to describe strangeness enhancement \cite{4r}, $J/\psi$ suppression \cite{5r} and fixed $p_T$ suppression (jet quenching) \cite{6r}. The parameters of the model ($p_T$-shift due to the interaction of the produced particle with the medium and the cross-section of this interaction) are determined from the data on fixed $p_T$-suppression. Elliptic flow is then due to the different absorption of the produced particle for different values of its azimuthal angle -- due to the different path length inside the overlap region of the colliding nuclei, associated to each azimuthal angle.

The purpose of the present paper is to examine in a critical way the so-called radial flow, i.e. the increase of the inverse slope $T$ with the mass of the produced particle -- which has been observed in nucleus-nucleus collisions both at CERN 
and RHIC \cite {8r} energies and is absent in $pp$ collisions. RHIC results for identified particles~: $\pi^{\pm}$, $K^{\pm}$, $p$ and $\overline{p}$ are presented in Fig.~1 in different centrality bins \cite{8r}. The increase of $T$ with the mass was derived in hydrodynamic models \cite{9r} where~:
\beq
\label{1e}
T = T_0 + m <u_t>^2 \ .
\eeq

\noi $T_0$ is the freeze-out temperature and $<u_t>^2$ measures the strength of the radial transverse flow. The fits \cite{8r} to the data using eq. (\ref{1e}) are given by the dashed lines in Fig.~1. This fit provides support to the hydrodynamics picture. In particular the data are consistent with a value of the freeze out temperature $T_0$ independent of centrality. However, although the slope $<u_t>^2$ increases with centrality, it is worth noticing that the most peripheral bin (60 - 92~\%) corresponds to an average number of participating nucleons of 14. Thus, according to the hydrodynamic interpretation a perfect fluid is already produced at this comparatively low density. This naturally leads us to ask whether the origin of eq. (\ref{1e}) is a genuine final state collective effect (which should be already present for $N_{part} = 14$) or it is contained in the initial state (Cronin effect, shadowing, etc.).\par

The latter is supported by the following remark. Apart from the increase of the inverse slope $T$ with the mass described by eq. (\ref{1e}), we see in the data that $T$ increases with centrality. This increase is larger when the mass increases. Now, the RHIC data \cite{10r,11r} on fixed $p_T$ suppression indicate that, at fixed centrality, the suppression increases with $p_T$ in the low $p_T$ region ($p_T < 2$~GeV) we are interested in this paper. Indeed, it is widely recognized that, due to the combined effects of decreasing shadowing and of the Cronin effect, the initial value of the ratio
\beq
\label{2e}
R_{Au\ Au} = dN^{Au\ Au}/dy (b, p_T)/N_{coll} dN^{pp}/dy
\eeq

\noi in absence of final state interactions increases rapidly with $p_T$. At $p_T \sim 2$~GeV all calculations of the Cronin effect lead to an initial value of $R_{Au\ Au}$ close to unity \cite{12r}. On the contrary, the final value of $R_{Au\ Au}$ at $p_T = 2$~GeV is substantially lower than unity \cite{10r,11r}. Thus, there is a sizable suppression at $p_T = 2$~GeV, whereas at $p_T = 0$ the data agree with model calculations \cite{3r} without final state interaction -- indicating that there is no suppression at $p_T = 0$.

If this suppression results from the interaction with the medium, it produces a decrease of $T$ at each centrality. Obviously this decrease will be larger for increasing centrality. Therefore, the final state interaction with the medium has an effect which is exactly the opposite of the one seen in the data of Fig.~1 -- providing a strong indication that the observed phenomenon is already present in the initial condition. 

The remaining part of this paper is devoted to a quantitative study of this point. The final result is given in Fig.~1 where the full signs and full lines show our results for the ``radial flow'' in the initial condition.

In refs. \cite{3r,6r} the fixed $p_T$ suppression of a pion has been described in the framework of a final state interaction model. Due to its interaction with the medium the pions suffer a $p_T$-shift, $\delta p_T$. This produces a loss at a given $p_T$. There is also a gain due to pions originally produced at $p_T + \delta p_T$. Due to the fall-off of the $p_T$ distribution, the loss is larger than the gain, resulting in a net pion suppression. The loss and gain differential equation can be solved analytically. The solution gives the survival probability of the pion \cite{3r}. The same equation describes the survival probability of  a particle of type $i$
\beq
\label{3e}
S_i(y,p_t,b) = \exp \left \{ - \sigma_i \rho (b,y)\left [ 1 - {N_i(p_T + \delta p_T) \over N_i (p_T)}(b)\right ] \ln \left ( \rho (b,y)/\rho_{pp}(y)\right ) \right \} \ . 
\eeq

\noi Here $N_i (p_T)$ denotes the $p_T$-distribution of the inclusive production of particle $i$. With $\delta p_T \to \infty$, the quantity inside the square bracket is equal to unity and there is no gain term. With $\delta p_T = 0$ the loss and gain terms are identical and $S_i = 1$. $\rho (b,y)$ is the density of the medium~:  $\rho (b,y) = dN^{AA}/dy(b)/G(b)$, where $dN^{AA}/dy (b)$ is computed in the Dual Parton Model \cite{13r}, including shadowing corrections, and $G(b)$ is an effective transverse area computed using Wood-Saxon profile functions. All details can be found in \cite{3r}. $\rho_{pp}(y)$ denotes the $pp$ density. Within conditions specified in \cite{3r} the argument of the logarithm is equal to the ratio of freeze-out time to initial time. As stated above we shall restrict ourselves to the low $p_T$ range used in the determination of the inverse slope ($p_T < 2$~GeV). 

The pion suppression data can be described \cite{3r,6r} using $\sigma = 1.3$~mb and $\delta p_T = p_T^{1.5}/(20$~GeV$^{1/2}$). In principle $\sigma$ can depend on the type of particle produced. However, as it has been argued in \cite{3r,6r}, the final state interactions described by eq. (\ref{3e}) take place largely at the partonic level -- in which case all $\sigma$'s should have similar values. Moreover, it has been shown in the second paper of \cite{3r} that the mass dependence of $v_2$ for $\pi$, $K$ and $p$ can be described by assuming that the cross-section for the three types of particles are the same. (The decrease of $v_2$ with the mass is due to the fact that the $p_T$-dependence gets broader with increasing mass. In this way the size of the gain term, relative to the loss one, increases. This produces a decrease of both the suppression and $v_2$ with increasing mass). Thus we shall use in the following the same value $\sigma = 1.3$~mb for the three types of particles. It is, however, worth noticing that if the value of $\sigma$ increases with the mass, the mass dependence of the slope $<u_t>^2$ in eq. (\ref{1e}) in the initial condition, resulting from our calculation, will increase (see below). Thus, by taking the same $\sigma$ for pions, kaons and protons we minimize the strength of the ``radial flow'' present in the initial condition.

Let us now turn to the only remaining unknown quantity in eq. (\ref{3e}) namely the one inside the square bracket. It is well known that the $p_T$ dependence of the inclusive spectra in the low $p_T$ region can be described by an exponential in the variable $m_T - m$ (where $m_T$ is the transverse mass) \cite{8r} 
\beq
\label{4e}
{d^2 N \over 2 \pi m_T dm_Tdy}(b) = {1 \over 2 \pi T(b) (T(b) + m_0))} A(b) \exp \left ( - {m_T - m_0 \over T(b)}\right )
\eeq

\noi where $T(b)$ is the inverse slope. The results of the PHENIX collaboration for identified $\pi^{\pm}$, $K^{\pm}$, $p$ and $\overline{p}$ in three rapidity intervals are shown in Fig.~2. The fits using eq. (\ref{4e}) are given in Fig.~3 of ref. \cite{8r}. The resulting values of $T(b)$ are also given in \cite{8r} and shown in Fig.~1 of the present paper. 

All considerations in this paper, as well as the experimental results in Figs.~1 and 2 refer to the mid-rapidity region $|y| < 0.35$. We use eq. (\ref{4e}) in order to determine the quantity in the square brackets in eq. (\ref{3e}). In this way all quantities in the survival probability are determined and we can proceed to the calculation of the initial $p_T$-distribution before the final state interaction has been switched on. These initial distributions are simply obtained by dividing, for each value of $p_T$ and $b$, the experimental result by the corresponding survival probability eq. (\ref{3e}), i.e. 
\beq
\label{5e}
{d^2N_i^{Initial} \over 2 \pi m_T dm_T dy}(b) = {d^2N_i \over 2 \pi m_T dm_T dy} (b)/S_i (y, p_T, b)
\eeq

The results for the initial $p_T$ distribution obtained in this way are shown in Fig.~2 (full lines).  The resulting values of the initial inverse slope $T^{Initial}$ for the various types of particle, in three centrality bins, is given in Fig.~1. We see that the inverse slopes of the initial $p_T$-distributions are larger than the measured ones and their differences increase with centrality. We also see that the ``radial flow'', i.e. the increase of $T$ with mass is already present in the initial condition. Actually, the slope $<u_t>^2$ of the mass dependence, eq. (\ref{1e}), is larger in the initial condition than in the data.

Note that in the calculations presented above we have computed the quantity inside brackets in eq. (\ref{3e}) using expression (\ref{4e}) with the measured values of $T$. It is easy to check that using instead the values $T^{Initial}$, corresponding to the initial $p_T$ distributions, the results are quite similar. Actually, one should solve the gain and loss differential equations numerically (instead of using their analytic solution, eq. (\ref{3e})). In this way the values of $T$ would evolve from $T^{Initial}$ at initial time $\tau_0$ to the measured values of $T$ at freeze-out time $\tau_f$. Obviously, the result will be intermediate between the ones obtained using the analytic solution, eq. (\ref{3e}), with either the initial or the final values of $T$. Since the difference between these two results is only of a few percent it is not worth solving the equation numerically. 

Our consideration have been restricted to light particles $\pi^{\pm}$, $K^{\pm}$, $p$ and $\overline{p}$. However, it is natural to expect that the increase of $T^{initial}$ with mass will saturate for larger masses. Indeed, the shadowing corrections decrease with increasing mass and the Cronin effect is expected to saturate for large masses. In our approach if such a saturation exists in the initial condition it will also exist in the final one, i.e. after final state interaction. Thus, particles such as $\Omega$, $J/\psi$, ... are expected to show no ``radial flow'' while we expect for them a non-vanishing value of $v_2$ \cite{3r}.

Another obvious prediction of our approach is a substantial ``radial flow'' in $dAu$ collisions. 

In conclusion, experimental data on $Au$ $Au$ collisions at $\sqrt{s} = 200$~GeV (as well as SPS energy ones) show that the inverse slope $T$ increases with centrality and this increase is larger with increasing mass of the trigger. In a final state interaction model that reproduces the observed fixed $p_T$ suppression, the final state interaction produces a {\it decrease} of the inverse slope and this decrease is larger with increasing mass of the trigger. This indicates that the so-called ``radial flow'' is an effect which is present in the initial condition -- rather than a collective phenomenon resulting from the interaction with the hot medium.

\begin{figure}
\centering\leavevmode
\ \ \ \
\vskip 1.5cm
\epsfxsize=6in\epsfysize=6in\epsffile{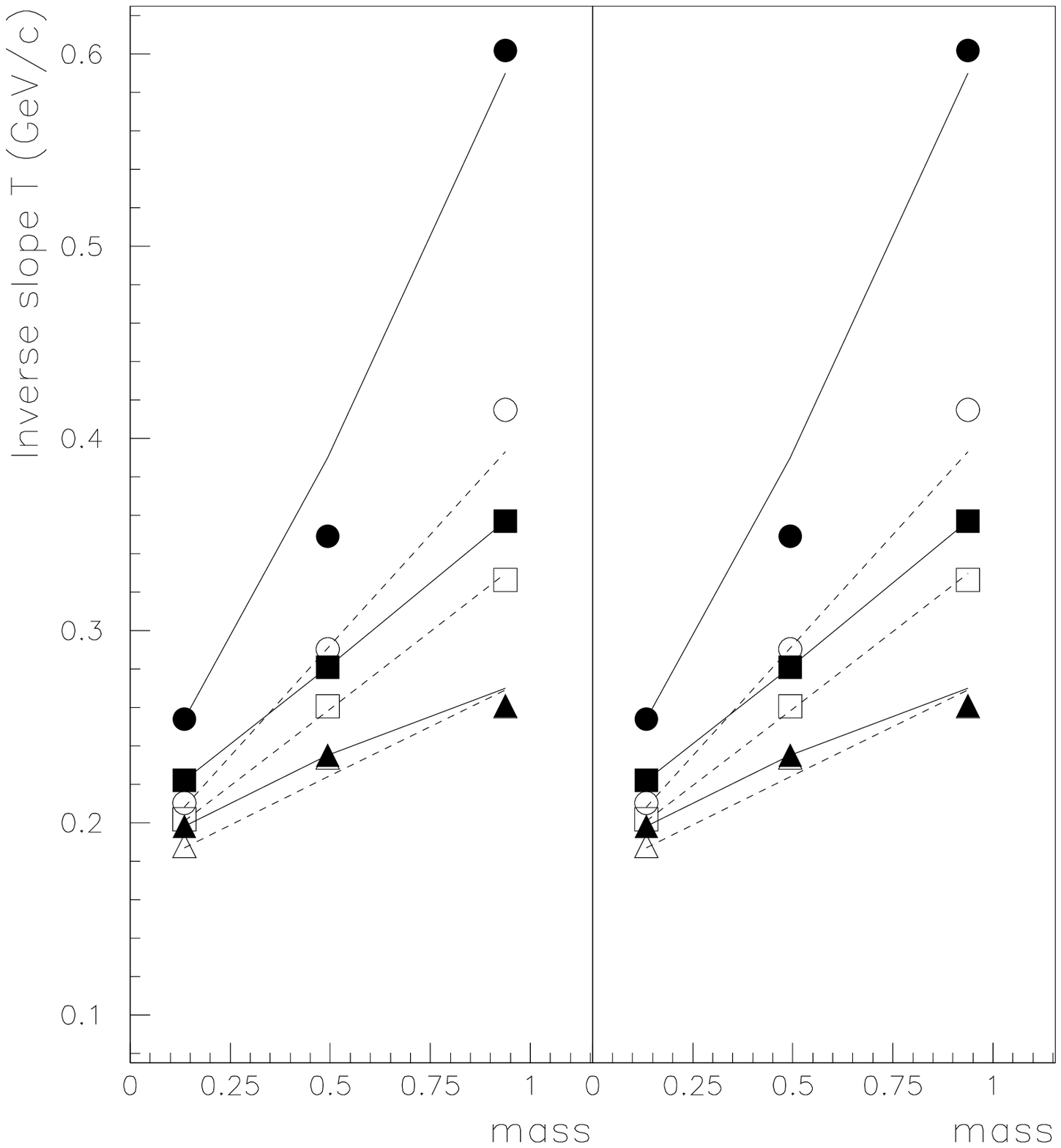}
\caption{Inverse slope $T$ versus mass for $\pi^+$, $K^+$ and $p$ (left) and $\pi^-$, $K^-$ and $\overline{p}$ (right) in three centrality bins for $Au$ $Au$ collisions at $\sqrt{s} = 200$~GeV/c. The data points (empty signs) and the dashed lines are from ref. {\protect\cite{8r}}. The full signs and full lines are the corresponding values for the initial condition, calculated from eq. ({\protect\ref{5e}}). The circles correspond to central (0-5~\%), the squares to mid-central (40-50~\%) and the triangles to peripheral (60-92~\%) collisions.}
\end{figure}
 
\newpage
 
\begin{figure}
\centering\leavevmode
\ \ \ \
\vskip 1.5cm
\epsfxsize=6in\epsfysize=6in\epsffile{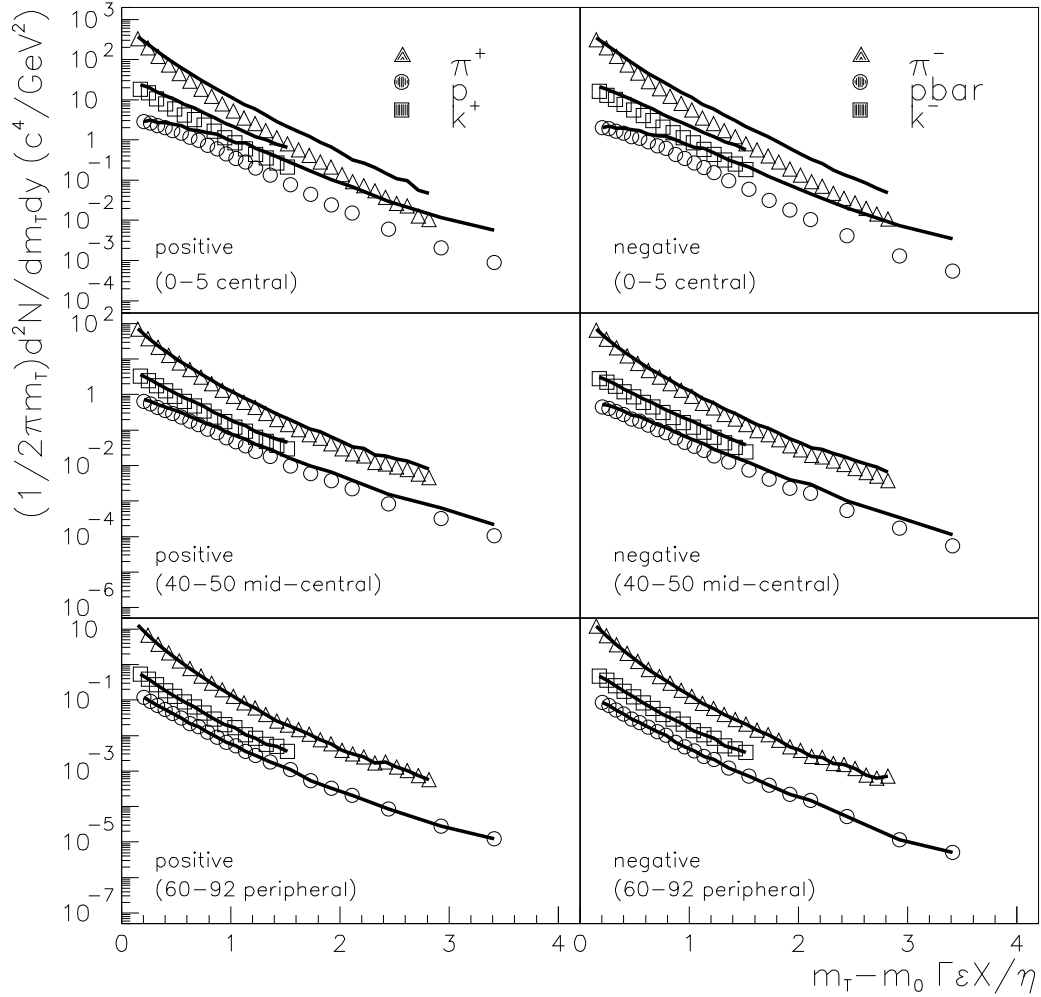}
\caption{Final (data points) and initial (lines) transverse mass distributions for $\pi^+$, $K^+$ and $p$ (left) and $\pi^-$, $K^-$ and $\overline{p}$ (right) in three centrality bins for $Au$ $Au$ collisions at $\sqrt{s} = 200$~GeV. The data are from {\protect\cite{8r}}. The lines are calculated using eq. ({\protect\ref{5e}}) and correspond to the initial condition. As expected, the effect of the final state interaction is important for central collisions and negligibly small for the most peripheral ones.}
\end{figure}
\end{document}